\documentclass[5p,twocolumn,times,number]{elsarticle}
\usepackage{lineno,hyperref}

\journal{Journal of Nuclear Instrument and Methods A}

\biboptions{sort&compress}
\begin{document}

\begin{frontmatter}

\title{The liquid argon detector and measurement of SiPM array at liquid argon temperature}

\renewcommand{\thefootnote}{\fnsymbol{footnote}}
\author{
C.~Guo$^{a,c}\footnote{Corresponding author: Tel:~+86-1088236256. E-mail address: guocong@ihep.ac.cn (C.~Guo).}$,
M.Y.~Guan$^{a}$,
X.L.~Sun$^{b}$,
W.X.~Xiong$^{a,c}$,
P.~Zhang$^{a,c}$,
C.G.~Yang$^{a,c}$,
Y.T.~Wei$^{a,c}$
Y.Y.~Gan$^{a,c}$,
Q.~Zhao$^{a,c}$
}

\address{
${^a}$Key Laboratory of Particle Astrophysics, Institute of High Energy Physics, Chinese Academy of Science,Beijing, China\\
${^b}$ State Key Laboratory of Particle Detection and Electronics, Institute of High Energy Physics, Beijing, China\\
${^c}$ School of Physics, University of Chinese Academy of Science, Beijing, China \\
}

\begin{abstract}

Particle detectors based on liquid argon (LAr) have recently become recognized as an extremely attractive technology for the direct detection of dark matter as well as the measurement of coherent elastic neutrino-nucleus scattering (CE$\nu$NS). The Chinese argon group at Institute of High Energy Physics has been studying the LAr detector technology and a LAr detector has been operating steadily. A program of using a dual phase LAr detector to measure the CE$\nu$NS at Taishang Nuclear Power Plant has been proposed and the R\&D work is ongoing. Considering the requirements of ultra-low radio-purity and high photon collection efficiency, SiPMs will be a good choice and will be used in the detector. In this proceeding, an introduction of the LAr detector and the measurement results of SiPM array at LAr temperature will be presented.
\end{abstract}


\end{frontmatter}


\section{Introduction}

The COHERENT collaboration has recently achieved the first measurement of elastic neutrino-nucleus scattering (CE$\nu$NS) by using a CsI(Na) crystal detector to detect the neutrinos from the spallation neutron source at Oak Ridge National Laboratory (ORNL). The CE$\nu$NS, which is predicted in 1974 as a sequence of neutral weak current~\cite{PRD1974,JETP1974}, is the dominant neutrino interaction for neutrinos of energy lower than 100 MeV. The characteristic dependence on the square of neutrons (N$^2$) reflects on the coherent sum of the weak charge carried by the neutrons and is sensitive to the nuclear form factor~\cite{JPG2009,PRL2018,PRD2019,PRC2012,JHEP2019,PRD2019-99}, which can be seen in the differential cross section of a spin-zero nucleus (Eq.~\ref{Eq.crosssection}):

\begin{equation}
\frac{d\sigma}{dT} = \frac{{G_F}^2M}{2\pi}\left[2-\frac{2T}{E_\nu}+\left(\frac{T}{E_\nu}\right)^2-\frac{MT}{{E_\nu}^2}\right] \frac{{Q_W}^2}{4}F^2(Q^2)
\label{Eq.crosssection}
\end{equation}

where $\sigma$ is the cross section, $T$ is the recoil energy, $M$ is the mass of the nucleus, $Q_W$ = N-Z(1-4sin$^2$$\theta_W$) is the weak charge with weak mixing angle $\theta_W$. CE$\nu$NS is sensitive to physics beyond the Standard Model (SM)~\cite{PRD2006,PRD2016,PRD2018}.

To detect the CE$\nu$NS, three basic requirements have to be met: 1. $\sim$~few keV nuclear recoil energy threshold, 2. intense sources or large target mass, 3. a low background detector. Liquid argon (LAr) is a very good target candidate for CE$\nu$NS detection because it has a high light yield, 40 photons/keVee (electron equivalent energy deposition), providing a low threshold. Further more, by using the amplification effects of the S2 signal in a dual phase LAr detector, the threshold can be down to sub-keV~\cite{DS-lowmass}. On the other hand, the depleted argon, which has been discovered and used in DS-50 experiment by DarkSide collaboration~\cite{DS-UAr}, made it more competitive in low background experiments. Apart from that, the two components of LAr scintillation light, the fast component with 7~ns decay time and slow component with 1.6~$\mu$s decay time, provides powerful pulse shape discrimination (PSD) capabilities to separate nuclear recoils (NR) and electronic recoils (ER)~\cite{PRC2008,AP2008}, which means that the $\gamma$ background coming from the surrounding environment can be further rejected.

Silicon photo-multipliers (SiPM), which were originally developed in Russia in the mid-1980s~\cite{NIMA2006}, have known a fast development in the last few years as a possible alternative to traditional photo-multiplier tubes (PMT). Compared with the traditional PMT, the SiPM has numerous advantages such as low bias voltage ($<$ 100V), high photon detection efficiency (PDE) for visible and near infrared photons, excellent single photon response, fast rise time ($<$ 1~ns), low power consumption, insensitivity to magnetic fields. Moreover, the SiPM has a comparable gain to PMT. All these advantages make SiPM attractive and competitive in the field of particle physics.

A project of measuring the CE$\nu$NS for reactor neutrinos with a LAr detector has been proposed, and the R$\&$D work is ongoing at Institute of High Energy Physics (IHEP). In this proceeding, an introduction of the LAr detector and a preliminary result of testing SiPM array at liquid argon temperature will be presented.

\section{The liquid argon detector at IHEP}
\subsection{Detector setup}
\begin{figure}[htb]
\centering
\includegraphics[height=7.5cm]{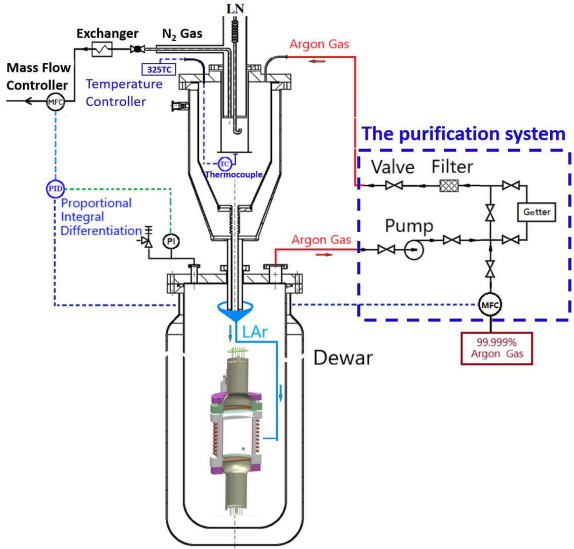}
\caption{Scheme of the LAr detector. }
\label{fig.LAr}
\end{figure}

Fig.~\ref{fig.LAr} shows the scheme of the LAr detector at IHEP. A liquid nitrogen (LN) cooling system is used to liquify the high purity (99.999\%) argon gas and the liquid argon is contained in the dewar. Inside the dewar, a  polytetrafluoroethylene (PTFE) sleeve with a dimension of 8~cm in diameter and 10.5~cm in height is used to contain 0.74~kg liquid argon. Two 3 inch PMTs (R11065) are placed at the top and bottom of the sleeve to detect the scintillation light of the LAr. A layer of enhanced specular reflector film (ESR) is placed on the surface of the PTFE sleeve to enhance the reflectivity. Two pieces of silica glass are placed at the top and bottom of the sleeve to generate a uniform electric filed together with 9 copper rings which are uniformly placed along the vertical axis. An etched stainless steel grid is placed 5~mm below the liquid-gas surface to generate the extraction field. All the inner surfaces, including the ESR and PMT windows are coated with 1,1,4,4-tetraphenyl-1,3-butadiene (TPB) to shift the 128~nm argon scintillation light to 420~nm. The detail information of the detector can be found in Ref.~\cite{Weixing}.

\subsection{Detector calibration}
Two radioactive sources, $^{83m}$Kr and $^{22}$Na, have been used to calibrate the LAr detector.

\begin{figure}[htb]
\centering
\includegraphics[height=6cm]{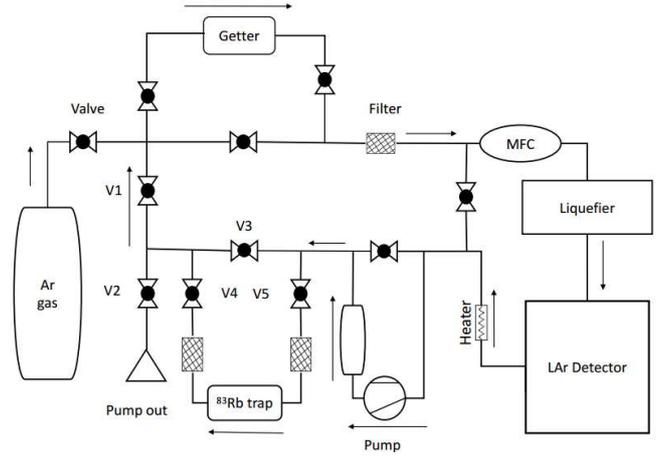}
\caption{The gas handling system of the LAr detector. While loading $^{83m}$Kr, valves V4 and V5 are open and V3 is closed. }
\label{fig.gassystem}
\end{figure}

The $^{83m}$Kr is released by a $^{83}$Rb source, which was produced at the institute of Modern Physics, Chinese Academy of Science. The $^{83m}$Kr decays with a half-life of 1.83$\pm$0.02~h via two electromagnetic transitions and release 32.1~keV and 9.4~keV $\gamma$ rays.  The $^{83}$Rb, which has a half-life of 86.2 days, is infused in a 0.45~g of zeolite, which is located in a gas cell called $^{83}$Rb trap in the gas-handling system. As is shown in Fig.~\ref{fig.gassystem}, $^{83m}$Kr can be introduced into the detector through the circulation system without compromising the purity of the liquid argon.

\begin{figure}[htb]
\centering
\includegraphics[height=6cm]{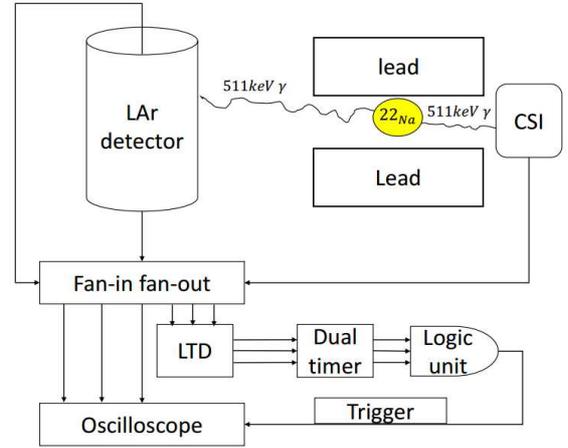}
\caption{The logical diagram for $^{22}$Na calibration. }
\label{fig.Na22}
\end{figure}

Before introducing the $^{83m}$Kr into the detector, a $^{22}$Na $\gamma$ source is used to calibrate the detector. $^{22}$Na emits a 1275~keV $\gamma$ ray and a positron, which almost annihilates immediately and produce two back-to-back 511~keV $\gamma$ rays. As is shown in Fig.~\ref{fig.Na22}, one 511~keV $\gamma$ ray aims at the center of the active LAr target and is collimated by a 13~cm thick lead collimator with a hole in 10~mm diameter, the other 511 $\gamma$ ray aims at a CsI crystal, which is coupled with a R5902 PMT for photon detection. To suppress the background from environmental radioactivities, the coincident signal from the LAr and CsI are used to form the trigger.

\begin{figure}[htb]
\centering
\includegraphics[height=5.3cm]{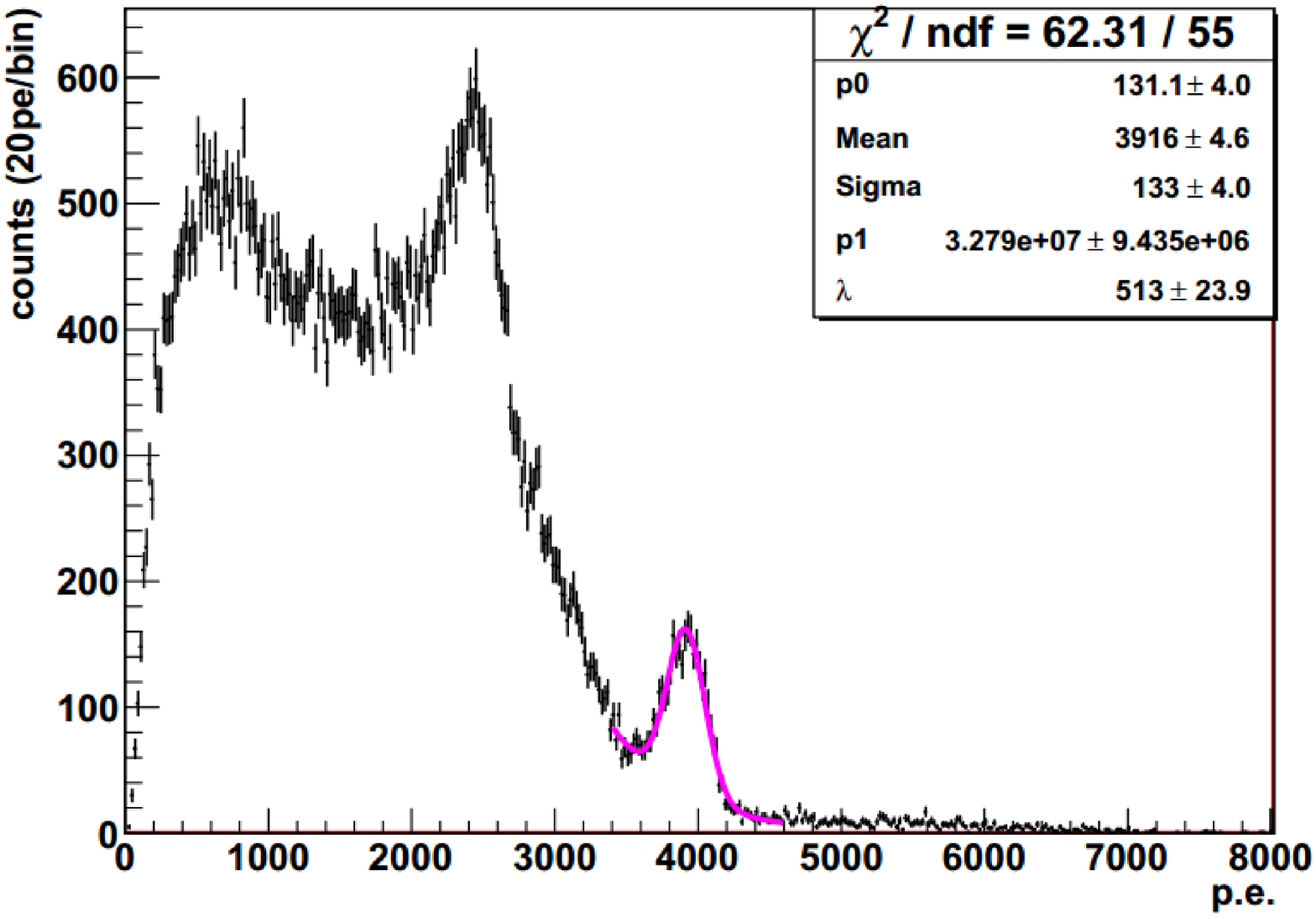}
\caption{The energy spectrum of $^{22}$Na and the purple line is the fitting result of the full energy peak of the 511~keV $\gamma$. }
\label{fig.spectrum_Na}
\end{figure}

\begin{figure}[htb]
\centering
\includegraphics[height=5.3cm]{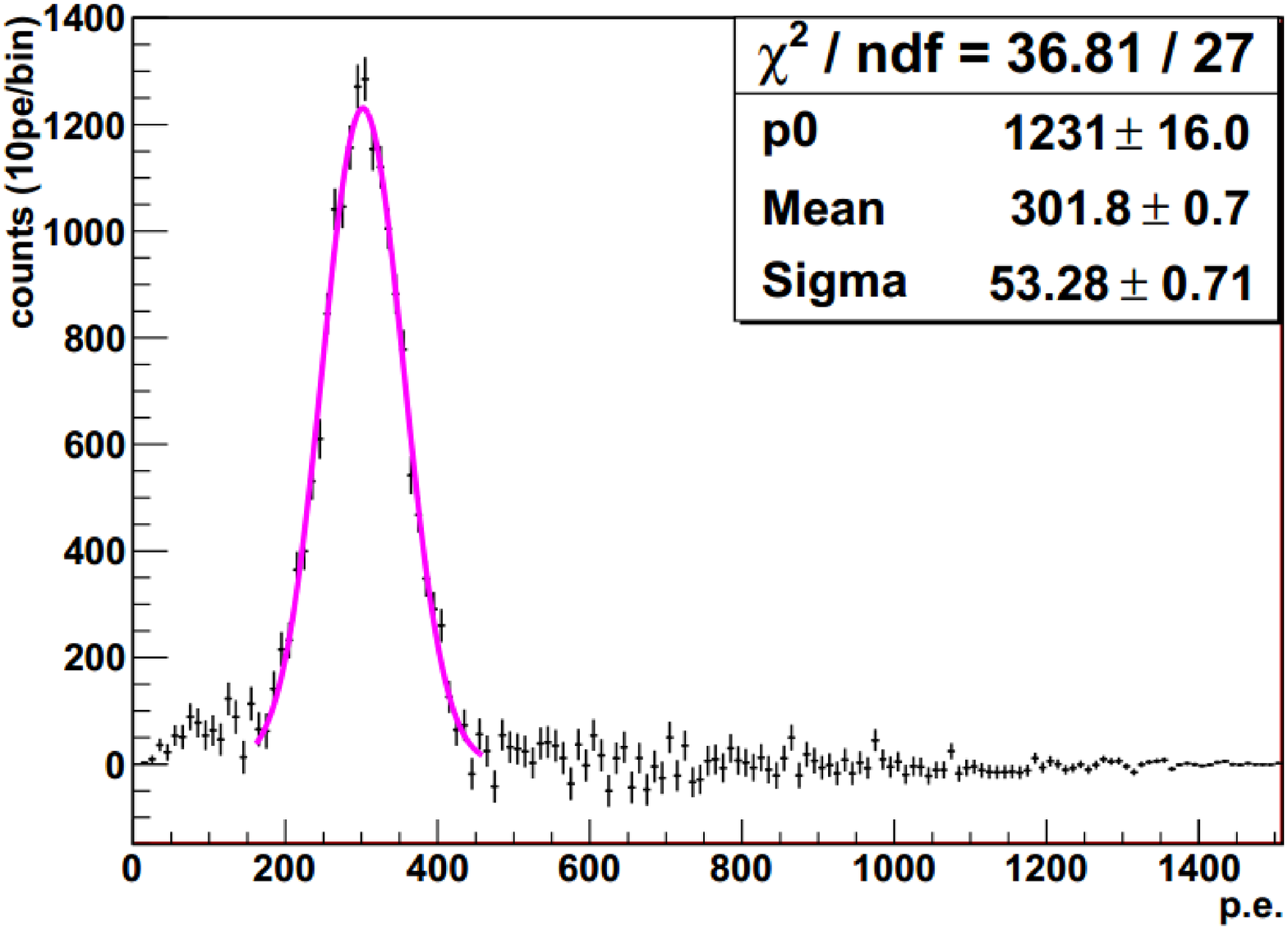}
\caption{The energy spectrum of $^{83m}$Kr and the purple line is the fitting result of the full energy peak of the 41.5~keV $\gamma$. }
\label{fig.spectrum_Kr}
\end{figure}

The calibration is carried out at null drift field and the energy spectrum for $^{22}$Na and $^{83m}$Kr are shown in Fig.~\ref{fig.spectrum_Na} and Fig.~\ref{fig.spectrum_Kr}. For $^{22}$Na, the spectrum is fitted with a function which consists of a Gaussian and an exponential and a light yield of 7.66 $\pm$ 0.01 photoelectrons/keV (p.e./keV) with an energy resolution of 3.6\% are obtained. While for $^{83m}$Kr, because the two $\gamma$s come from the cascade decay of the same nucleon, only a peak of the total energy can be observed in the energy spectrum. A gaussian function with a least squares fitting method is used to fit the spectrum and a light yield of 7.28 $\pm$ 0.02 p.e./keV with a energy resolution of 17.6\% are obtained. The light yield difference between $^{22}$Na and $^{83m}$Kr is because LAr has a greater stopping power for lower energy $\gamma$s or electrons~\cite{ester}.

When a particle interacts with an argon atom, the excited atoms (excitons) and the ionized atoms (ions) will be produced. The excitons decay via the formation of an excited dimer and produce photons while the ions undergo the recombination process to form an excited dimer. In a dual phase LAr detector, the existing drifting field will help some of the ions to escape the recombination and resulting a decrease in the light yield~\cite{PRB1979}. The variation of the light yield with different drift field from 0 to 200~V/cm have been measured with $^{22}$Na. As is shown in Tab.~\ref{table1}, a decrease of 21.4\% of the light yield is observed at 200~V/cm drifting field compared with null field.

\begin{table}[h]
\begin{center}
\begin{tabular}{|c|c|c|c|}
\hline
E[V/cm] & $\mu_p$[p.e.] & $\sigma_p$[p.e.] & LY[p.e./keV] \\ \hline
0    & 3916 $\pm$ 4.6  & 133.0  &   7.66 $\pm$ 0.01 \\ \hline
50   & 3607 $\pm$ 5.7  & 142.7  &   7.06 $\pm$ 0.01  \\ \hline
100  & 3390 $\pm$ 6.1  & 147.5  &   6.63 $\pm$ 0.01  \\ \hline
150  & 3220 $\pm$ 5.9  & 149.5  &   6.30 $\pm$ 0.02  \\ \hline
200  & 3078 $\pm$ 6.9  & 147.3  &   6.02 $\pm$ 0.02  \\ \hline
\end{tabular}
\caption{The mean value ($\mu_p$), resolution ($\sigma_p$) and light yield (LY) obtained from $^{22}$Na 511~keV full energy peak at different drift field (E). The errors are statistical only.}
\label{table1}
\end{center}
\end{table}

\section{Tests of SiPM array}
The silicon photomultiplier (SiPM) is a single photon sensitive light sensor that combines performance characteristics that exceed of a PMT, with the practical advantages of a solid state sensor. A SiPM consists of a parallel array of many avalanche photodiodes (APDs), where each APD micro-pixel independently works in limited Geiger mode with an applied voltage of a few volts above the breakdown voltage (V$_{bd}$). When a photoelectron is produced, a Geiger avalanche will be induced. The avalanche is passively quenched by a resistor integrated to each pixel. The output charge from a single pixel is independent of the number of the produced photoelectrons within the pixel. The number of fired pixels is proportional to the number of injected photons if it is smaller compared to the total number of the pixels.
\subsection{Experimental setup}
\begin{figure}[htb]
\centering
\includegraphics[height=5cm]{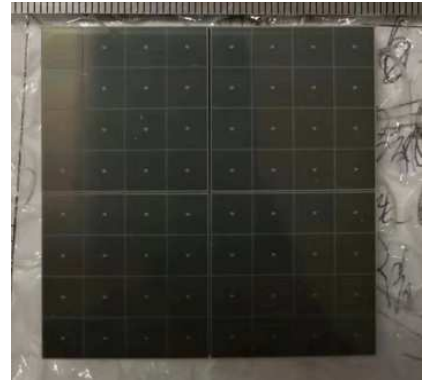}
\caption{The picture of the SiPM array used in the test. }
\label{fig.SiPM}
\end{figure}

\begin{figure}[htb]
\centering
\includegraphics[height=5 cm]{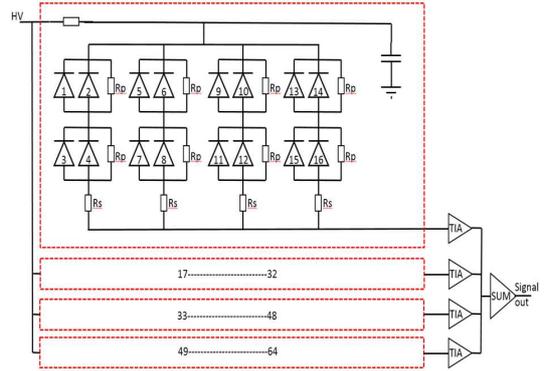}
\caption{The schematic diagram of the readout board. }
\label{fig.readout}
\end{figure}

A 5~cm $\times$ 5~cm SiPM array, which consists of 64 pieces of Hamamastu S14160 SiPM chips, has been tested. Fig.~\ref{fig.SiPM} shows the picture of the SiPM array and Fig.~\ref{fig.readout} shows the schematic diagram of the readout board.

\begin{figure}[htb]
\centering
\includegraphics[height=5 cm]{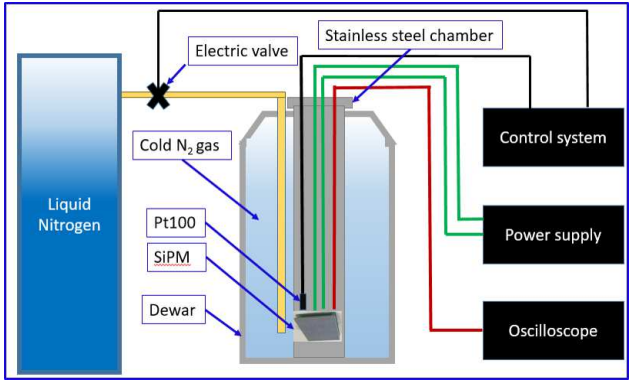}
\caption{The schematic diagram of the measurement system.}
\label{fig.testsystem}
\end{figure}

During the measurements, the SiPM array is placed in a stainless steel chamber which is put in a dewar. As is shown in Fig.~\ref{fig.testsystem}, a fully automatic nitrogen filling system and a temperature control system are used to fulfill the temperature adjustment from 77K to room temperature. A Pt100 sensor is placed inside the dewar for temperature monitoring as well as serving as an input of the control system. During the measurement, the temperature difference inside the stainless steel chamber is within $\pm$1~K. A direct current electrical source is served as the power supply system to supply the bias for the SiPM and power up the pre-amplifiers. The power consumption of the readout board is $\sim$250~mW. A LeCroy 610Zi oscilloscope is used for pulse record.

\subsection{Gain and break down voltage}
\begin{figure}[htb]
\centering
\includegraphics[height=5 cm]{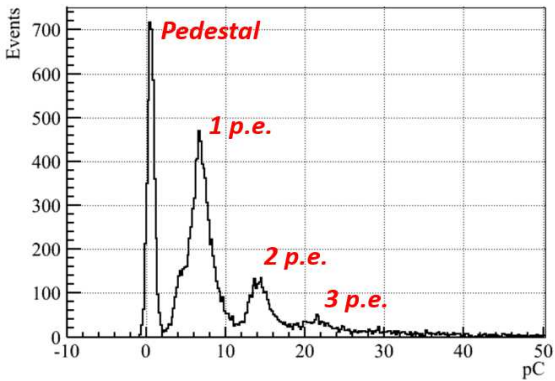}
\caption{The energy spectrum of dark noise for the SiPM at 87K.}
\label{fig.spectrum_sipm}
\end{figure}

\begin{figure}[htb]
\centering
\includegraphics[height=5 cm]{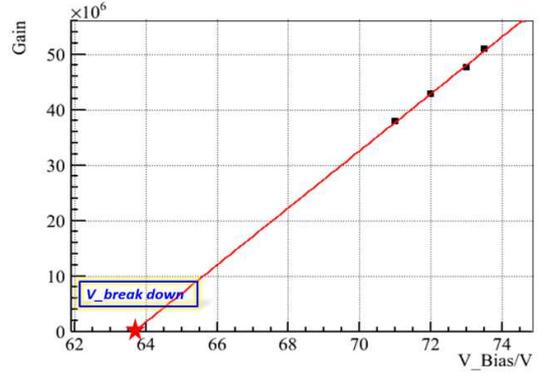}
\caption{Determination of V$_{bd}$ from the gain-voltage extrapolation at 87K.}
\label{fig.breakdown}
\end{figure}

\begin{figure}[htb]
\centering
\includegraphics[height=4.8 cm]{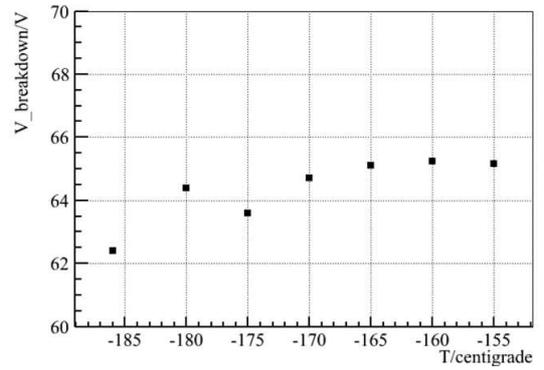}
\caption{Measured break down voltage at different temperature.}
\label{fig.bdt}
\end{figure}

As is shown in Fig.~\ref{fig.spectrum_sipm}, the peaks of different photoelectrons are well separated, and the gain of the SiPM can be easily gotten by measuring the charge corresponding to one p.e. and dividing it by the electron charge, namely:

\begin{equation}
Gain = \frac{Q}{e} = \frac{C(V-V_{bd})}{e}\equiv\frac{C{V_{over}}}{e}
\label{Eq.gain}
\end{equation}

where $Q$ is the charge of one p.e., $V$ is the bias supplied on the SiPM, $C$ is the capacitance, $V_{bd}$ is the breakdown voltage, $V_{over}$ is the over voltage and $e$ is the charge of an electron. As is shown in Eq.~\ref{Eq.gain}, the gain is proportional to $V_{over}$, thus the $V_{bd}$ can be derived by linearly extrapolating the gain-voltage relation to the point where gain becomes zero. As is shown in Fig.~\ref{fig.breakdown}, the breakdown voltage for the tested SiPM is 63.4~V at 87~K. The gain is dependent on the over voltage and the break down voltage is dependent on the temperature, which can be seen from Fig.~\ref{fig.bdt}.

\subsection{Optical Crosstalk}
\begin{figure}[htb]
\centering
\includegraphics[height=4.8 cm]{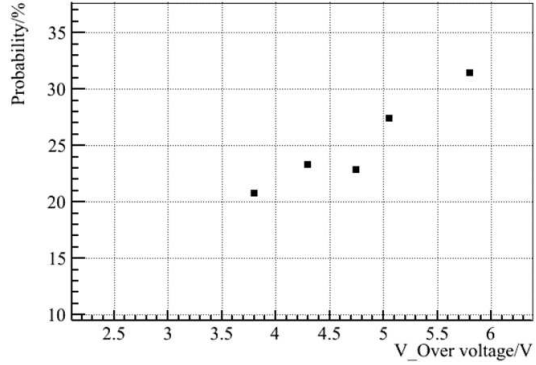}
\caption{The measured optical cross talk probability for different over voltage at 87K .}
\label{fig.crosstalk}
\end{figure}

The optical crosstalk between different $\mu$cell is one component of the SiPM noise. When undergoing avalanche, carriers near the junction emit photons as they are accelerated by the high electric field. These photons tend to be in the near infrared region and can travel substantial distances through the device. The crosstalk probability is the probability that an avalanching $\mu$cell will initiate an avalanche in a second $\mu$cell. The process happens instantaneously and as a consequence, single photons may generate signals equivalent to a 2, 3 or higher photoelectron events. The optical cross talk probability can be estimated by the ratio of the counting rate at a threshold of second p.e. to the counting rate at single p.e.. Fig.~\ref{fig.crosstalk} shows the measured optical cross talk probability for different over voltage at LAr temperature.

\subsection{Dark counting rate}
\begin{figure}[htb]
\centering
\includegraphics[height=4.8 cm]{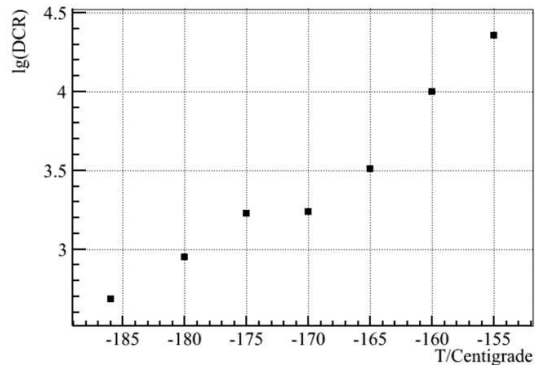}
\caption{The measured DCR for the SiPM at different temperature. The threshold is set to $\sim$0.5~p.e. and the over voltage is $\sim$5~V. }
\label{fig.DCR}
\end{figure}

The dark count rate (DCR) represents the number of avalanches per second generated by carriers traversing the depleted region when the SiPM is kept in dark condition. DCR is an important parameter because it sets a limit on the minimum detectable signal of a given device. The DCR includes primary and secondary signals. The primary dark pulses are mainly due to carriers thermally generated in the depletion regions of the $\mu$cells. The secondary signals are related to the after pulses and optical cross talk. As is shown in Fig.~\ref{fig.DCR}, the dark count rate is 482~Hz for the array, namely $\sim$0.2~Hz/mm$^2$. When measuring the DCR, the threshold is set to 0.5~p.e. level and the over voltage is $\sim$5~V.

\section{Summary and prospect}

The R$\&$D of using a liquid argon detector to measure CE$\nu$NS for reactor neutrinos is ongoing in IHEP. A dual phase liquid argon detector is running well and a 5 $\times$5  cm$^2$ SiPM array has been developed and measured at liquid argon temperature. The preliminary results show that the Hamamatsu S14160 SiPM is a good candidate for photon detection for LAr detector. A detail measurement of the SiPM array in liquid argon will be conducted recently.

\section{Acknowledgements}
This work is supported by Ministry of Science and Technology of the People's Republic of China (2016YFA0400304).


\end{document}